\documentclass[useAMS,usenatbib]{mn2e}
\usepackage{graphicx}
\usepackage{amssymb}
\usepackage{epsfig}
\usepackage{color}

\title[On the deceleration of jets]%
{On the deceleration of relativistic jets in active galactic nuclei I: Radiation drag}
\author[V.~S.~Beskin and A.~V.~Chernoglazov]
{V.~S.~Beskin$^{1,2}$\thanks{E-mail: beskin@lpi.ru (VSB)} and A.~V.~Chernoglazov$^{2}$ \\
$^{1}$ Lebedev Physical Institute, Russian Academy of Sciences,
Leninsky prospekt 53, Moscow 119991, Russia\\
$^{2}$ Moscow Institute of Physics and Technology, Institutsky per.~9,
Dolgoprudny 141700, Russia\\
}

\begin{document}

\date{Accepted \dots Received \dots; in original form \dots}

\pagerange{\pageref{firstpage}--\pageref{lastpage}} \pubyear{2016}

\maketitle

\label{firstpage}

\begin{abstract}
Deceleration of relativistic jets from active galactic nuclei (AGNs) detected recently
by MOJAVE team is discussed in connection with the interaction of the jet material with
the external photon field. Appropriate energy density of the isotropic photon field
which is necessary to decelerate jets is determined. It is shown that the disturbances
of the electric potential and magnetic surfaces play important role in general dynamics
of particle deceleration.
\end{abstract}


\section{Introduction}
\label{s:intro}

Recent progress in VLBI observations of relativistic jets outflowing from active galactic
nuclei~\citep{AGN1, AGN2, AGN3, AGN4} gives us new information concerning their physical
characteristics and dynamics. In particular, rather effective deceleration of the jet material
on the scale more than 50--100 pc was recently detected by MOJAVE team~\citep{MOJAVE12}. We
consider one possible explanation of such a deceleration connecting with the interaction of
the jet with the external photon field. Both radiation drag and particle loading will be
considered in detail on the ground of standard MHD approach, the first mechanism below and
the second one in the accompany paper~\citep{paper2}.

Remember that it is the magneto-hydrodynamical (MHD) model that is now developed intensively
in connection with the theory of relativistic jets outflowing from a rotating supermassive
($M \sim 10^{8}\hbox{--}10^{9}\,M_{\odot}$) black holes, which are thought
as a 'central engine' in active galactic nuclei and quasars~\citep{BBR84,TPM86}. In particular,
it is the MHD model that is now the most popular in connection with the problem of the origin
and stability of jets. Moreover, within last several years additional observational
confirmations were found in favor of the MHD model such as the presence of the $e^+e^-$
plasma~\citep{Rey96, Hir98} as well as the toroidal magnetic field~\citep{Gab92}.
Finally, recent numerical simulations~\citep{num1, num2, num3} demonstrate their very nice
agreement with MHD analytical asymptotic solutions.

On the other hand, the density of photons in the vicinity of the central engine is high
enough. This implies that the photon field may change drastically the characteristics of
the ideal MHD outflow. For example, they may result in the particle loading, i.e., extensive
$e^+e^-$ pair creation~\citep{Sw}, their acceleration by action of the radiation drag force
for small enough particle energies as well as the deceleration of high energetic
particles~\citep{Sikora1}. In other words, in the self-consistent consideration the
interaction of the magnetically dominated flow with the external photon field is to
be taken into account.

Unfortunately, many years these two processes, i.e., MHD acceleration and the action of
external photons, was developed separately. Only in the paper by~\citet{LBC92} the first
analytical step was done to combine them together. In particular, it was demonstrated
how general equations can be integrated for conical geometry (which is impossible in
general case). On the other hand, the consideration was produced in the given poloidal
magnetic field. But under this assumption the fast magnetosonic surface (for cold flow)
locates at infinity~\citep{Mich69, KFO76, Lery98}. As a result, it was impossible to
analyze the radiation drag effect in the vicinity of the fast magnetosonic surface and
the properties of the supersonic flow outside this surface.

Self-consistent disturbance of magnetic surfaces was included into consideration
by~\citet{BZS-04} for high enough particle energy (when the radiation pressure is
ineffective in particle acceleration). It was demonstrated that for magnetically
dominated flow the drag force actually does not change the particle energy diminishing
only the total energy flux. It was shown as well that the disturbance of magnetic surfaces
becomes large only if the drag force changes significantly the total energy flux. Finally,
recently~\citet{Russo1, Russo2} have considered the drag action on the magnetised
outflow in gamma-busters where the radiation pressure can play the leading role in particle
acceleration.

As to partile loading, several aspects of this process were considered
by~\citet{Sw, Lyu03,  DAKK03, SP06}. Even if the electron-positron pairs
are created at rest (and, hence, they do not change the total energy and angular
momentum flux), increasing of the particle flux inevitably decreases the
mean particle energy. As a result, the particle loading can be considered as a
rather effective mechanism of the deceleration of the jet bulk motion as well.

The main goal of this paper is to determine more carefully the photon drag action
on the cylindrical magnetically dominated outflow. As a zero approximation (i.e.,
without radiation drag and particle loading) we use well-known analytical solution
for cylindrical magnetically dominated MHD outflow~\citep{IP94, Beskinbook}. As we
are interesting in the region far enough from the 'central engine', in what follows
we consider the simple isotropic model of the radiation field (i.e. for energy density
$U = U_{\rm iso} =$ const). Actually, our goal is just in evaluating $U_{\rm iso}$
which are necessary to explain the observable deceleration of jets on the scale
$50$--$100$ pc.

The paper is organized as follows. At first in Sect.~\ref{sub:problem} we discuss the
very necessity to use two-fluid MHD approximation for highly magnetized winds and jets
in the presence of the external photon field. In Sect.~\ref{s:RD}, starting from the basic
two-fluid MHD equations we demonstrate how the drag force redistributing the electric
charges results in the appearance of longitudinal electric field. It gives us the possibility
to determine the change of particle energy. The beam damping resulting from particle loading
is discussed in the accompany paper~\citep{paper2}. Finally, in  Sect.~\ref{sect:diss}
the main results of our consideration including astrophysical applications are formulated.


\section{A Problem}
\label{sub:problem}

At first, let us formulate the main unsolved problem we are going to discuss. Up to now, both
analytically~\citep{Mich69, GJ2, HN, A&C, BKR, BN-06}
and numerically~\citep{Komissarov-94, Ustyu95, BTs99, num1, Tch08, Tch09, Bucc, num2, num3},
the properties of highly magnetized winds and jets were mainly described within MHD
approximation. Only recently the first steps were done using PIC numerical
simulation~\citep{Sironi1, MmSAI}, but these explorations are still in the very beginning.

It is convenient for us to introduce just now the main dimensionless parameters describing
ideal MHD flow, namely, the particle multiplicity $\lambda$, the magnetization parameter
$\sigma_{\rm M}$, and the compactness parameter $l_{a}$. First, to describe the flow
number density one can introduce so-called particle multiplicity $\lambda$
\begin{equation}
\lambda = \frac{n^{({\rm lab})}}{n_{\rm GJ}},
\label{lam}
\end{equation}
where $n_{\rm GJ}=|\rho_{\rm GJ}|/e$ and $\rho_{\rm GJ} = \Omega_{0} B_{0}/(2\pi c)$
is the~\citet{GJ1} charge density, i.e., the minimum charge density required for the
screening of the longitudinal electric field in the flow. Here $B_{0}$ is the poloidal
magnetic field in a jet and $\Omega_{0}$ is the central engine angular velocity. As was
shown by~\citet{NBKZ}, for active galactic nuclei the multiplication parameter can be
very large: $\lambda \sim 10^{11}$--$10^{13}$.

Next,~\citet{Mich69} magnetization parameter $\sigma_{\rm M}$ shows by how much the
electromagnetic energy flux near the central engine can exceed the particle energy
flux. The value $\sigma_{\rm M}$ corresponds to the maximal bulk Lorentz factor of
the plasma that can be reached in the case where all the electromagnetic energy
is transferred to the particle flow. In other words, $\sigma_{\rm M}$ is the maximum
Lorentz factor that can be achieved in the magnetized wind. For cylindrical flow
under consideration one can determine $\sigma_{\rm M}$ as
\begin{equation}
\sigma_{\rm M} = \frac{\Omega_{0} e B_0 r_{\rm jet}^2}{4\lambda m_{\rm e} c^3},
\label{c_sigma}
\end{equation}
where  $r_{\rm jet}$ is its transverse dimension of a jet.

The convenience of these two parameters stems from the fact that their product depends on
the total energy losses $W_{\rm tot}$ only and, hence, can be determined from observations.
Indeed, as was shown by~\citet{Beskin-10},
\begin{equation}
\lambda \sigma_{\rm M}
\sim \left(\frac{W_{\rm tot}}{W_{\rm A}}\right)^{1/2},
\label{newsigma}
\end{equation}
where $W_{\rm A} = m_{\rm e}^{2}c^{5}/e^{2} \approx 10^{17}$ erg/s. This value corresponds
to minimum energy losses of a 'central engine' which can accelerate particles up to
relativistic energies. Hence, we obtain $\lambda \sigma_{\rm M} \sim 10^{14}$ for ordinary
jets from AGN. Another representation of the product $\lambda \sigma_{\rm M}$ is
\begin{equation}
\lambda \sigma_{\rm M}
\sim \frac{e E_{r} r_{\rm jet}}{m_{\rm e} c^2},
\label{newsigmaa}
\end{equation}
where $E_{r} \sim (\Omega_{0} r_{\rm jet}/c) B_{0}$. As we see, this value corresponds
to the total potential drop across the jet.

Finally, the compactness parameter
\begin{equation}
l_{a} = \frac{\sigma_{\rm T} U_{\rm iso} R}{m_{\rm e}c^2}
\label{compact}
\end{equation}
is in fact the optical depth by Thomson cross section $\sigma_{\rm T}$ at a distance $R$
in the photon field with energy density $U_{\rm iso}$. Below, it will be important for us
that the parameter $l_{a}$  provide an upper limit of particle energy in the acceleration
region. On the other hand, a large $l_{a}$  is necessary for effective particle production.

It is necessary to stress that in this paper we consider only leptonic model of the relativistic
jets. For this reason we normalize all the values on electron mass $m_{\rm e}$. This approach is
reasonable for very central parts of a jet connecting by magnetic field lines with the black hole
horizon (and, hence, loaded by secondary $e^+e^-$ plasma generated by  photon-photon conversion).  It
is in this region the numerical simulations mentioned above demonstrate regular magnetic field
and energy flux.  As to periphery part of a jet connecting with accreting disk, the special
consideration including reconnection is necessary. This is beyond the scope of the present consideration.

Returning now to one-fluid MHD approach, it is necessary to stress that it has serious
restriction. Indeed, well-known freezing-in condition ${\bf E} + {\bf v} \times {\bf B}/c = 0$
results in two consequences
\begin{eqnarray}
E_{\parallel} & = & 0, \\
E_{\perp} & < & B,
\end{eqnarray}
namely, zero longitudinal electric field and smallness of the perpendicular electric field
in comparison with magnetic one. \citet{Ferraro} isorotation law, i.e., the conservation of
so-called field angular velocity $\Omega_{\rm F}$ (see below) along magnetic tubes is the
mathematical formulation of this property. As a result, very large potential difference between
the center and the boundary of a jet takes place up to the very end of a flow where the jet meets
the external media (lobes in AGNs, HH objects in YSO, stellar wind in close TeV binaries).

On the other hand, it is clearly impossible to describe the interaction of external media
with highly magnetized flow without including this potential drop into consideration. Indeed,
neglecting $E_{\perp}$ we do not include into consideration the role of the Poynting flux
which is the main actor of our play. As a result, during such an interaction the domains
with nonzero longitudinal electric field or with $E > B$ are to appear resulting in very
effective particle acceleration~\citep{Beskin-10}. Nevertheless, up to now the role of the
Poynting flux during the interaction with external media was considered only indirectly, say,
by adding large enough toroidal magnetic field which energy density is similar to that of
magnetized flow~\citep{BHeid1, BHeid2, DelaCita}. Remember that general properties of the
MHD shock containing arbitrary Poynting flux were already formulated more than ten years
ago~\citep{Double}.

Effective particle acceleration can takes place even without external media. As was already
demonstrated many years ago~\citep{BGI93, BR2000}, if there is some restriction on the
longitudinal electric current circulating in the magnetosphere of radio pulsar, in the
vicinity of the light cylinder $R_{\rm L} = c/\Omega$ the region with $E > B$ appears. As a
result, in the narrow region $\Delta r \sim R_{\rm L}/\lambda$ the very effective particle
acceleration is to take place up to the bulk Lorentz-factor $\Gamma \sim \sigma_{\rm M}$.
It is interesting that just such a sudden acceleration was recently supposed to explain
pulse TeV radiation from Crab pulsar~\citep{nature}{\footnote{The title of this paper is
'Abrupt acceleration of a cold ultrarelativistic wind from the Crab pulsar'.}}. Moreover,
recent PIC modelling of the axisymmetric pulsar magnetosphere~\citep{Cerutti} also
demonstrates very effective particle acceleration near the light cylinder up to
$\gamma \sim \sigma_{\rm M}$.

Here it is necessary to stress one very important point. Not only one-fluid, but
even two-fluid MHD approximation is not sufficient to describe the interaction of
the highly magnetized flow with external media. As was shown by~\citet{BGI93, BR2000},
effective particle acceleration in the domain with $E > B$ inevitably accompanied
by vanishing of the radial velocity. This implies many-fluid regime which cannot
be described analytically. The same concerns another dissipative processes, say,
the magnetic reconnection which also discussed, mainly phenomenologically~\citep{rec1,
rec2, rec3, rec4, rec5, LevGlo16} and numerically~\citep{BarKom, DelZanna, TaMa},
in connection with the energy release in the highly magnetized flow.

In this paper we are not going to discuss the very interaction of a jet with external media, but
try to evaluate the role of the external photon field in hydrodynamical retardation of a jet. In
this case two-fluid approximation allows us to include into consideration self-consistently the
longitudinal electric field and the disturbance of magnetic surfaces. As a result, one-fluid
validity condition will be formulated.

\section{Radiation drag}
\label{s:RD}

\subsection{Qualitative Consideration}
\label{sub:qualit}

\begin{figure}
\includegraphics[scale=0.47]{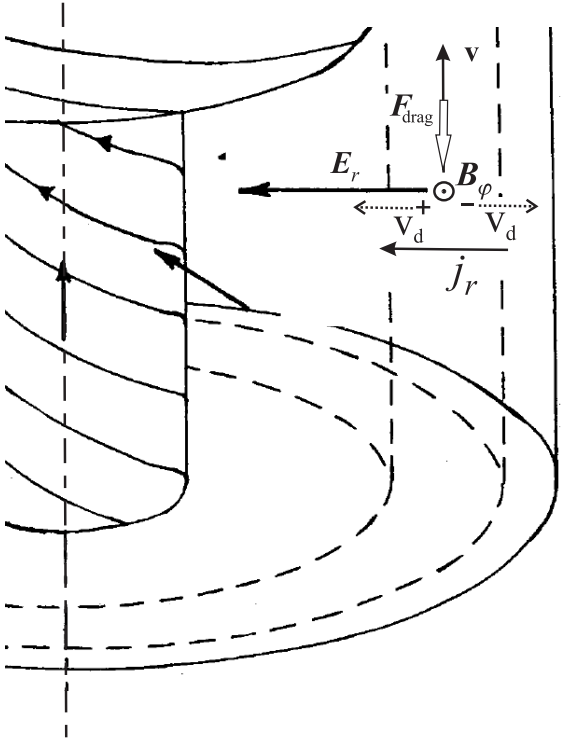}
\caption{
Drag force ${\bf F}_{\rm drag}$ results in the appearance of the radial drift current,
redistribution of the electric charges, diminishing of the radial electric field and, finally,
damping of the Poynting flux. The particle energy remains actually constant as the negative
work of the drag itself actually equal to energy gain resulting from particle intersection
of equi-potential surfaces.
}
\label{fig01}
\end{figure}

At first, let us consider interaction of the magnetically dominated jet with the
isotropic photon field qualitatively. Without the drag far enough from the rotation
axis the particle motion along the jet corresponds to electric drift in radial electric
$E_r$ and toroidal magnetic $B_{\varphi} \gg B_z$ fields~\citep{Tch08, Beskinbook}.
It is clear that the drag force ${\bf F}_{\rm drag}$ directed along the jet results
in the radial drift of electrons and positrons in opposite directions (see Fig.~\ref{fig01}).
The appropriate electric current can be evaluated as
\begin{equation}
j_r \sim \lambda \rho_{\rm GJ} V_{\rm d},
\label{jdr}
\end{equation}
where
\begin{equation}
V_{\rm d} \sim c\frac{F_{\rm drag}}{e B_{\varphi}}
\label{vdr}
\end{equation}
is the drift velocity. Such a current is to diminish the toroidal magnetic field $B_{\varphi}$.
Simultaneously, redistribution of charges is to diminish the radial electric field $E_{r}$. Both
these processes result in reducing of the Poynting vector flux.

As in the magneticallym dominated jet \mbox{$E_{r} \approx B_{\varphi}$,} one can write down
the energy equation for the time-independent flow $\nabla \, {\bf S} = - {\bf j\,E}$ as
\begin{equation}
\frac{c}{4 \pi} \, \frac{{\rm d}B_{\varphi}^2}{{\rm d}z} \approx - j_r B_{\varphi}.
\label{Energy}
\end{equation}
Using now relation (\ref{newsigma}) and evaluations
$B_{\varphi}/B_z \sim r_{\rm jet}/R_{\rm L}$ and
$W_{\rm tot} \sim (c/4 \pi) B_{\varphi}^2 r_{\rm jet}^2$,
we finally obtain for the characteristic retardation scale $L_{\rm dr}$
\begin{equation}
L_{\rm dr} \sim \sigma_{\rm M} \, \frac{m_{\rm e}c^2}{F_{\rm drag}}.
\label{scale}
\end{equation}
The same evaluation can be directly obtained from the continuity equation
$\nabla \, {\bf j} = 0$
\begin{equation}
\frac{j_{r}}{r_{\rm jet}} \sim \frac{j_{\parallel}}{L_{\rm dr}},
\label{divj}
\end{equation}
where $j_{\parallel} \approx \rho_{\rm GJ} c$.

As we see, the work $A_{\rm dr} = F_{\rm drag} L_{\rm dr}$  of the drag force $F_{\rm drag}$
on the scale $L_{\rm dr}$ (resulting in IC photons release of the jet energy flux)
\begin{equation}
A_{\rm dr} \sim \sigma_{\rm M} m_{\rm e}c^2
\label{Adr}
\end{equation}
just equals to the particle energy corresponding to the total energy transfer from
the electromagnetic Poynting flux to plasma outflow~\citep{BR2000}. This implies
that our evaluation of the retardation scale is correct. But as this force acts
actually perpendicular to the large toroidal magnetic field
$B_{\varphi} \sim (\Omega_{0} r_{\rm jet}/c)B_{0}$, in the first approximation the
energy of particles remains constant. The point is that the energy loss
$-F_{\rm drag}v_{z}$ resulting from the drag force will be fully compensated by the
energy $e E_{r} V_{r}$ gaining by particles due to their radial drift motion along
radial electric field. Another words, the particle energy remains actually constant
as the negative work of the drag itself actually equal to energy gain resulting
from particle intersection of equi-potential surfaces.

Thus, as was firstly demonstrated by~\citet{BZS-04}, the drag force acting on
plasma particles in the highly magnetized wind results in not the diminishing of
particle energy but the diminishing of the Poynting flux
as both the toroidal magnetic and radial electric field decrease along the jet.
As to particle deceleration, this process appears in the second order approximation
when we have to include into consideration the diminishing of integrals of motion.

\subsection{Cylindrical Flow}
\label{sub:cylinder}

\subsubsection{Basic Equations}
\label{ssub:basic_RD}

In this section we consider the interaction of the cylindrical magnetically dominated jet with
the isotropic photon field quantitatively. To have the possibility to analyze this process
analytically, some simplifications will be used. At first, as was already stressed, we
discuss leptonic model of relativistic jets. Second, below we consider pure cylindrical jet.
This assumption is more serious than one can imagine at first glance. The point is that
cylindrical geometry implies infinite curvature of the poloidal magnetic field. In this case
there is well-known asymptotic behavior for particle Lorentz factors in the magnetically
dominated flow $\Gamma \approx  \Omega r_{\perp}/c$ which will be used in what follows.

Remember that for finite curvature radius $R_{\rm c}$ of the poloidal magnetic field another
asymptotic solution $\Gamma \approx (R_{\rm c}/r_{\perp})^{1/2} $ is possible. As was shown
by~\citet{BZS-04}, see also~\citet{LevGlo16}, under some conditions the photon drag increase
 the curvature radius $R_{\rm c}$ resulting in bulk acceleration of plasma
particles. In addition, intrinsic instabilities of cylindrical jets~\citep{benford, hardee2,
A&C, l99, NalBeg12, TchBro16} also can change drastically the dynamics of interaction of photon
field with magnetically dominated outflow. These processes are beyond the scope of our present
consideration.

Finally, in what follows we consider magnetically dominated jet, i.e., the jet which does
not reach their terminal Lorentz factor $\Gamma = \sigma_{\rm M}$. It is not clear that
the flow remains highly magnetized up to the distances 10-100 pc from the central engine
under consideration. Nevertheless, this case is more interesting from physical point of
view as it gives us the possibility to include into consideration the interaction of the
photon field with Poynting flux. In Sect. 4 some astrophysical applications connecting
with FRI--FRII classification will be given.

Thus, following~\citet{BZS-04}, we write down the set of time-independent Maxwell equations
and two-fluid equations of motion for electron-positron plasma:
\begin{eqnarray}
\nabla \, {\bf E}  =  4\pi \rho_{\rm e},~~~~~~~
\nabla \times {\bf E} =0,\label{1_1} \\
\nabla \, {\bf B}  =  0,~~~~~~~
\nabla \times {\bf B} = \frac{4 \pi}{c}{\bf j}, \label{1_2}\\
({\bf v}^{\pm}\nabla){\bf p}^{\pm}  =
e\left( {\bf E}+ \frac{{\bf v}^{\pm}}{c}\times{\bf B}\right)
+ {\bf F}_{\rm drag}^{\pm}.
\label{1_3}
\end{eqnarray}
Here $\bf E$ and $\bf B$ are the electric and magnetic fields,  $\rho_{\rm e}$ and $\bf j$
are the charge and current densities, and ${\bf v}^{\pm}$ and ${\bf p}^{\pm}$  are the speed
and momentum of particles. Finally, ${\bf F}_{\rm drag}$ is the radiation drag force. For
isotropic photon field~\citep{Blumenthal&Gould, Rybicki&Lightman}
\begin{equation}
{\bf F}_{\rm drag}^{\pm} = -\frac{4}{3}\frac{\bf v}{v}\sigma_TU_{\rm iso} \,
(\gamma^{\pm})^2,
\end{equation}
where $\gamma^{\pm}$ are the Lorentz-factor of particles.

As is well-known, in the axisymmetric case one can express the
electric and magnetic fields through three scalar functions, $\Psi(r_{\perp}, z)$,
$\Omega_{\rm F}(r_{\perp}, z)$, and $I(r_{\perp}, z)$
\begin{eqnarray}
{\bf B} & = & \frac{\nabla \Psi \times {\bf e}_{\varphi}}{2 \pi r_{\perp} }
- \frac{2 I(\Psi)}{c r_{\perp} }{\bf e}_{\varphi}, \\
{\bf E} & = & -\frac{\Omega_{\rm F}(\Psi)}{2 \pi c} \nabla \Psi.
\end{eqnarray}
Here $\Psi(r_{\perp}, z)$ is the magnetic flux, $I(\Psi)$ is the total electric
current within the same magnetic tube, and $\Omega_{\rm F}(\Psi)$ is the so-called
field angular velocity (more exactly, the angular velocity of plasma drifting in
electromagnetic fields).

For cylindrical outflow we have the following force-free solution of
the general equations (\ref{1_1})--(\ref{1_3})~\citep{IP94}
\begin{eqnarray}
4\pi I(\Psi) = 2 \, \Omega_{\rm F}(\Psi)\Psi
\label{Izero}
\end{eqnarray}
corresponding to homogeneous poloidal magnetic field
\begin{eqnarray}
B_{z}^{(0)} = B_0,
\end{eqnarray}
so that $\Psi^{(0)} = \pi B_0 r_{\perp} ^2$, i.e., it does not depend on coordinate $z$,
\begin{eqnarray}
B_{\varphi}^{(0)} & = & -\frac{2I}{cr_{\perp}}, \\
E_{r}^{(0)} & = & B^0_{\varphi},
\label{E&B}
\end{eqnarray}
and
\begin{eqnarray}
B_r^{(0)}=0,~~~ ~~B_{z}^{(0)} = B_0,~~~ E_{\varphi}^{(0)}=0,~~~
E_{z}^{(0)}=0.
\end{eqnarray}
It is important that this solution can be realised by massless particles moving along
the jet with the velocity equal to that of light
\begin{eqnarray}
v_z^{(0)}=c,~~~ v_{r}^{(0)}=0, ~~~ v_{\varphi}^{(0)}=0.
\end{eqnarray}
Moreover, this solution is true for arbitrary profile of the angular velocity
$\Omega_{\rm F}(\Psi)$. In particular, one can consider the most interesting
case $I(\Psi_{\rm jet}) = \Omega_{\rm F}(\Psi_{\rm jet}) = 0$, when the total
electric current flowing within the jet is equal to zero. For this reason, in
what follows we consider $\Omega_{\rm F}(r_{\perp} )$ as an arbitrary function.

As previously, in the cylindrical case we seek the first-order corrections for the
case $v \ne c$ in the following manner:
\begin{eqnarray}
n^{+} & = & \frac{\Omega_0 B_0}{2\pi c e}
\left[\lambda - K(r_{\perp})
+\eta^{+}(r_{\perp} , z)\right],
\label{c_cor1}\\
n^{-} & = & \frac{\Omega_0 B_0}{2\pi c e}
\left[\lambda + K(r_{\perp})
+\eta^{-}(r_{\perp}, z)\right], \label{c_cor2}\\
v_z^{\pm} & = &
c\left[1-\xi_z^{\pm}(r_{\perp} ,z)\right], \\
v_{r}^{\pm} & = & c\xi_{r}^{\pm}(r_{\perp} ,z), \\
v_{\varphi}^{\pm} & = & c\xi_{\varphi}^{\pm}(r_{\perp} ,z).
\label{c_cor4}
\end{eqnarray}
Here $\Omega_{0} =\Omega_{\rm F}(0)$ and again $\lambda = n_{\rm e}/n_{\rm GJ}$
(\ref{lam}) is the multiplicity parameter. As was already stressed, for active
galactic nuclei $\lambda \sim 10^{11}$--$10^{13}$.
Below, for simplicity, we consider $\lambda$ as a constant. Besides,
\begin{equation}
K(r_{\perp}) = \frac{1}{4r_{\perp} }\frac{{\rm d}}{{\rm d}r_{\perp} }
\left(r_{\perp} ^2\frac{\Omega_{\rm F}}{\Omega_0}\right)
\label{c_K}
\end{equation}
describes the charge density
\begin{equation}
\rho_{\rm e}^{0}(r_{\perp}) = -\frac{\Omega_{0}B_{0}}{\pi c} K(r_{\perp})
\label{rhoe}
\end{equation}
and current density $j_{z}^{0} = \rho_{\rm e}^{0}c$ transverse profiles. In particular,
$K(0) = 1/2$ and
\begin{equation}
\pi \int_{0}^{r_{\rm jet }} K(r') r' {\rm d} r' = 0,
\label{c_Kbis}
\end{equation}
so both the total charge and total longitudinal current in the jet vanish. Finally,
the disturbances of the electric potential $\Phi(r_{\perp},z)$ and magnetic flux
$\Psi(r_{\perp},z)$ can be written as
\begin{eqnarray}
\Phi(r_{\perp},z) & = &
\frac{B_0}{c} \left[\int_{0}^{r_{\perp}}\Omega_{\rm F}(r')r'{\rm d}r'
+\Omega_{0} r_{\perp}^2\delta(r_{\perp},z)\right],
\\
\Psi(r_{\perp},z) & = & \pi B_0 r_{\perp} ^2\left[1 + f(r_{\perp} ,z)\right].
\label{c_cor3}
\end{eqnarray}
It gives
\begin{eqnarray}
B_{r} & = & -\frac{1}{2}r_{\perp}  B_0
\frac{\partial f}{\partial z},
\label{c_cor5} \\
B_{\varphi} & = & -\frac{\Omega_0r_{\perp}}{c}B_0
\left[\frac{\Omega_{\rm F}}{\Omega_0} + \zeta(r_{\perp} ,z)\right],
\label{c_cor6} \\
B_{z} & = & B_0\left[1+\frac{1}{2r_{\perp}}
\frac{\partial}{\partial r_{\perp}}\left(r_{\perp} ^2
f\right)\right],
\label{c_cor7} \\
E_{r} & = &
- \frac{\Omega_0r_{\perp} }{c}B_0\left[\frac{\Omega_{\rm F}}{\Omega_0}
+ \frac{1}{r_{\perp} }\frac{\partial}{\partial r_{\perp} }(r_{\perp} ^2\delta)\right],
\label{c_cor8} \\
E_{z} & = & -\frac{\Omega_0r_{\perp} ^2}{c}B_0\frac{\partial\delta}{\partial z}.
\label{c_cor9}
\end{eqnarray}
As we see, the values $|\delta| \sim 1$ and $|f| \sim 1$ just correspond
to almost full dissipation of the Poynting flux.

Substituting now expressions (\ref{c_cor1})--(\ref{c_cor9}) into (\ref{1_1})--(\ref{1_3}),
we obtain to the first order approximation the following linear system of equations:
\begin{eqnarray}
-\frac{1}{r_{\perp} }\frac{\partial}{\partial r_{\perp} }(r_{\perp} ^2 \zeta)=
\nonumber \\
2(\eta^+-\eta^-)-2\left[\left(\lambda-K\right)\xi_z^+
-\left(\lambda+K\right)\xi_z^-\right],
\label{c_sys1} \\
2(\eta^+-\eta^-)+\frac{1}{r_{\perp} }\frac{\partial}{\partial
r_{\perp} }\left[r_{\perp} \frac{\partial}{\partial
r_{\perp} }\left(r_{\perp} ^2\delta\right)\right]+
r_{\perp} ^2\frac{\partial^2\delta}{\partial z^2}=0,
\label{c_sys2} \\
r_{\perp} \frac{\partial\zeta}{\partial z}
=2\left[\left(\lambda-K\right)\xi_{r}^+
-\left(\lambda+K\right)\xi_{r}^-\right],
\label{c_sys3}\\
- r_{\perp} ^2\frac{\partial^2 f}{\partial z^2}
- r_{\perp} \frac{\partial}{\partial r_{\perp} }\left[ \frac{1}{r_{\perp}}
\frac{\partial}{\partial r_{\perp} }\left( r_{\perp}^2 f\right) \right]=
\nonumber \\
4 \, \frac{\Omega_0r_{\perp} }{c}
\left[\left(\lambda-K\right)\xi_{\varphi}^+
-\left(\lambda+K\right)\xi_{\varphi}^-\right],
\label{c_sys4} \\
\frac{\partial}{\partial z}\left(\xi_{r}^+\gamma^+\right)=
-\xi_{r}^+F_{\rm d} (\gamma^{+})^2
\nonumber \\
+4 \, \frac{\lambda\sigma_{\rm M}}{r_{\rm jet}^2}\left[
-\frac{\partial}{\partial r_{\perp} }(r_{\perp} ^2\delta)+
r_{\perp} \zeta-r_{\perp} \frac{\Omega_{\rm F}}{\Omega_0}\xi_z^++
\frac{c}{\Omega_0}\xi_{\varphi}^+\right],\label{c_sys5}\\
\frac{\partial}{\partial z}\left(\xi_{r}^-\gamma^-\right)=
-\xi_{r}^-F_{\rm d}(\gamma^{-})^2
\nonumber \\
-4 \, \frac{\lambda\sigma_{\rm M}}{r_{\rm jet}^2}\left[
-\frac{\partial}{\partial r_{\perp} }(r_{\perp} ^2\delta)+
r_{\perp} \zeta-r_{\perp} \frac{\Omega_{\rm F}}{\Omega_0}\xi_z^-+
\frac{c}{\Omega_0}\xi_{\varphi}^-\right],
\label{c_sys6} \\
\frac{\partial}{\partial z}\left(\gamma^+\right)=
-F_{\rm d}(\gamma^{+})^2 + 4 \, \frac{\lambda\sigma_{\rm M}}{r_{\rm jet}^2}\left(
-r_{\perp} ^2\frac{\partial\delta}{\partial z}-
r_{\perp} \frac{\Omega_{\rm F}}{\Omega_0}\xi_{r}^+\right),
\label{c_sys7} \\
\frac{\partial}{\partial z}\left(\gamma^-\right)=
-F_{\rm d}(\gamma^{-})^2 - 4 \, \frac{\lambda\sigma_{\rm M}}{r_{\rm jet}^2}\left(
-r_{\perp} ^2\frac{\partial\delta}{\partial z}-
r_{\perp} \frac{\Omega_{\rm F}}{\Omega_0}\xi_{r}^-\right),
\label{c_sys8}\\
\frac{\partial}{\partial z}\left(\xi_{\varphi}^+\gamma^+\right)=
-\xi_{\varphi}^+F_{\rm d}(\gamma^{+})^2
\nonumber \\
+4 \, \frac{\lambda\sigma_{\rm M}}{r_{\rm jet}^2}\left(
-\frac{1}{2}\frac{cr_{\perp} }{\Omega_0}\frac{\partial f}
{\partial z}-\frac{c}{\Omega_0}\xi_{r}^+\right),
\label{c_sys9} \\
\frac{\partial}{\partial z}\left(\xi_{\varphi}^-\gamma^-\right)=
-\xi_{\varphi}^-F_{\rm d}(\gamma^{-})^2
\nonumber \\
-4 \, \frac{\lambda\sigma_{\rm M}}{r_{\rm jet}^2}\left(
-\frac{1}{2}\frac{cr_{\perp} }{\Omega_0}\frac{\partial f}
{\partial z}-\frac{c}{\Omega_0}\xi_{r}^-\right).
\label{c_sys10}
\end{eqnarray}
Here again $\sigma_{\rm M}$ (\ref{c_sigma}) is Michel magnetization parameter,
and $F_{\rm d} \approx l_{a}/R$ is the normalized radiation drag force
\begin{equation}
F_{\rm d} = \frac{4}{3} \, \frac{\sigma_{\rm T }U_{\rm iso}}{m_{\rm e} c^2}.
\label{F_K}
\end{equation}

\subsubsection{Zero MHD Approximation}
\label{ssub:zero}

As was already stressed, expression (\ref{Izero}) can be considered as a zero force-free
approximation describing cylindrical flow of massless particles. In the absence of a drag
force we can now find exact MHD solution describing pure cylindrical flow as well. Indeed,
as one can easily check, for
\begin{equation}
\left(\lambda-K\right)\xi_z^+=\left(\lambda+K\right)\xi_z^-,
\label{condition1}
\end{equation}
and
\begin{equation}
\xi_{\varphi}^{\pm} = x\xi_z^{\pm}
\label{condition2}
\end{equation}
the cylindrical flow with $\xi_{r}^{\pm}=0$, $\zeta = \delta = f =0$ results in
$\partial/\partial z =0$. Here and below we use dimensionless distance from the axis
$x_0 = \Omega_{0} r_{\perp}/c$,
and
\begin{equation}
x=\Omega_{\rm F}(r_{\perp})r_{\perp} /c.
\label{condition2_5}
\end{equation}

As we see, in this case it is necessary to introduce a small difference in velocity of particles
\begin{equation}
\xi_z^+-\xi_z^-=\frac{2K}{\lambda}\xi_z \sim \lambda^{-1}\xi_z,
\label{condition3}
\end{equation}
where $\xi_z=(\xi_z^++\xi_z^-)/2$ is the hydrodynamical velocity. It is not surprising
because equations (\ref{c_sys1})--(\ref{c_sys10}) now describe the flow in MHD (not
force-free) approximation. On the other hand, the mean particle energy is still the free
function.

Below we use the following notations
\begin{eqnarray}
\Gamma & = & \frac{\gamma^{+} + \gamma^{-}}{2}, ~~~G=\gamma^{+} - \gamma^{-}
\label{def1},
\\
P_{+} & = & \frac{\xi_z^+ + \xi_z^-}{2}, ~~~P_{-}=\xi_z^+ - \xi_z^-,
\label{def2}
\\
Q_{+} & = & \frac{\xi_{\varphi}^+ + \xi_{\varphi}^-}{2}, ~~~Q_{-}=\xi_{\varphi}^+ - \xi_{\varphi}^-,
\label{def3}
\end{eqnarray}
Finally, as a free function we choose
\begin{equation}
\Gamma^2  = \Gamma_{0}^2 + x^2,
\label{Gamma0}
\end{equation}
where $\Gamma_{0} \sim 1$ is the free parameter. Expression (\ref{Gamma0}) just corresponds
the well-known analytical asymptotic solution obtained in many papers, see~\citet{Beskinbook}
and references herein. Then, using relations (\ref{condition1})--(\ref{condition2}), one can
obtain
\begin{eqnarray}
Q_{\pm} & =& xP_{\pm}, \\
P_{-} & = & 2 \frac{K}{\lambda} P_{+},  \\
Q_{-} & = & 2 \frac{K}{\lambda} Q_{+}, \\
G & = & - \Gamma ^3 (1-x^2 P_+) P_{-},
\label{gamma-}
\end{eqnarray}
where
\begin{eqnarray}
P_{+} = \frac{1}{\Gamma (\Gamma + \sqrt{\Gamma ^2 - x^2})}.
\label{PP}
\end{eqnarray}
In the last expression we put square root into the denominator to avoid the subtraction
of two almost equal values $\Gamma$ and $\sqrt{\Gamma ^2 - x^2}$ in the numerator.

\subsection{Drift Approximation}
\label{Drift}

\subsubsection{Two-fluid effects}

Now we can use drag-free MHD solution (\ref{condition1})--(\ref{condition2}) and
(\ref{Gamma0})--(\ref{PP}) as a zero approximation, and evaluate the action of a
drag force finding small disturbances in the linear approximation. It is clear
that in this case all the disturbances including longitudinal electric field
$E_{\parallel}$ will be proportional to drag force $F_{\rm d}$. Thus, under some
conditions the electric force $e E_{\parallel}$ acting on the charged particle
could be larger than the retardation drag force $F_{\rm d}$. In this case one of
the species will be accelerated while another one will be decelerated more
efficiently than by action of the drag force only resulting in full stop at
some point. Thus, this condition corresponds to non-hydrodynamical regime.
For this reason the determination of the ratio $eE_{\parallel}/F_{\rm drag}$
is one of the main goal of our consideration.

Equations (\ref{c_sys1})--(\ref{c_sys10}) can be simplified in the drift approximation.
Indeed, well-known expression for drift velocity
\begin{equation}
{\bf V}_{\rm dr} = c \frac{(e{\bf E}+{\bf F}_{\rm drag}) \times {\bf B}}{e B^2}
\label{drift}
\end{equation}
fixes two velocity components perpendicular to the magnetic field ${\bf B}$.

It is necessary to remember that in the presence of the any force ${\bf F}$ having the
longitudinal component to the magnetic field the expression (\ref{drift}) is not
valid. On the other hand, moving into the reference frame in which the force ${\bf F}$
is parallel to the magnetic field, one can find that
\begin{equation}
\frac{|V_{\rm d}|}{c} =
\frac{1+\epsilon_{\perp}^2+\epsilon_{\parallel}^2-\sqrt{(1-\epsilon_{\perp}^2)^2
+ \epsilon_{\parallel}^2(2 + 2\epsilon_{\perp}^2+\epsilon_{\parallel}^2)}}{2 \epsilon_{\perp}},
\label{Dr}
\end{equation}
where $\epsilon_{\perp, \parallel} = F_{\perp, \parallel}/eB$, the direction of the
drift velocity remaining the same. As we see, the difference with the standard expression
(\ref{drift}) $|V_{\rm dr}|/c = \epsilon_{\perp}$ is proportional to $\epsilon_{\parallel}^2$.
Hence, in the linear approximation under consideration this correction can be neglected.

As a result, determining all the velocity components and substituting them into equations
of motion (\ref{c_sys7})--(\ref{c_sys8}), as it is shown in Appendix~\ref{A1}, one can obtain
\begin{eqnarray}
\frac{\partial \gamma^{\pm}}{\partial z} =
- \frac{(1-x^2P_{+})^2}{(1+x^2)} F_{\rm d} (\gamma^{\pm})^2
\nonumber\\
\mp \frac{4 \lambda \sigma_{\rm M}}{r_{\rm jet} ^2} \frac{(1-x^2 P_{+})}{(1+x^2)}
\left( -r_{\perp}^2 \frac{\partial \delta}{\partial z}
+r_{\perp}^2 \frac{\Omega_{\rm F}}{\Omega_0}
\frac{1}{2} \frac{\partial f}{\partial z} \right).
\label{gamma}
\end{eqnarray}
Expression (\ref{gamma}) (which is one of the key result of our consideration) can be also
obtained directly if we remember that general expression
\begin{equation}
\frac{{\rm d}{\cal E}}{{\rm d}t} = ({\bf F}_{\rm drag} + e{\bf E}){\bf v}
\label{edot}
\end{equation}
in the drift approximation (\ref{drift}) looks like
\begin{equation}
\frac{{\rm d}{\cal E}}{{\rm d}t} = (F_{\parallel} + eE_{\parallel})v_{\parallel}.
\label{edotpar}
\end{equation}
In other words, only longitudinal component of the force (and only longitudinal component
of the velocity) can change particle energy. Appearance of the factors $(1 + x^2)^{-1}$ and
 \begin{equation}
(1 -x^2 P_{+}) \approx \frac{\Gamma_{0}}{\Gamma} \ll 1
\label{corr}
\end{equation}
just result from this property.

As we see, together with the drag force (first term) always diminishing particle energy, Eqn.
(\ref{gamma}) contains the action of longitudinal electric field $E_{\parallel}$ having two sources.
In addition to disturbance of the electric potential $\delta$, longitudinal electric field
$E_{\parallel}$ is to appear due to disturbance of magnetic surfaces $f$. The last
term obviously vanishes if
\begin{equation}
\delta = \frac{1}{2}\frac{\Omega_{\rm F}}{\Omega _0}f,
\label{c_one2}
\end{equation}
i.e., if magnetic surfaces are equi-potential. Thus, one can conclude that self-consistent
analysis of the longitudinal electric field is to include into consideration not only the
disturbance of electric potential $\delta$, but the disturbance of the magnetic surfaces
$f$ as well.

As was already stressed, all linear disturbances are to be proportional to the drag force
$F_{\rm d}$. To determine these dependencies let us introduce the values
\begin{eqnarray}
g_{+} = \frac{\delta \gamma^{+} +\delta \gamma{-}}{2}, ~~~g_{-} = \delta \gamma^{+} -\delta \gamma{-},
\\
p_{+} = \frac{\delta \xi_z ^+ + \delta \xi_z ^-}{2}, ~~~p_{-} = \delta \xi_z ^+ - \delta \xi_z ^-,
\\
q_{+} = \frac{\delta \xi_{\varphi} ^+ + \delta \xi_{\varphi} ^-}{2},
~~~q_{-} = \delta \xi_{\varphi} ^+ - \delta \xi_{\varphi} ^-.
\end{eqnarray}
Substituting them into (\ref{c_sys1})--(\ref{c_sys10}) we obtain
\begin{eqnarray}
q_- = x p_-,
\label{d_sys1}\\
q_+ = x p_+ +\frac{1}{R_L}\frac{\partial }{\partial r_{\perp}}(r_{\perp}^2 \delta)- x_0 \zeta,
\label{d_sys2}\\
g_+ = - \Gamma ^3 p_+ + x\Gamma^3 P_+ q_+ + \frac{1}{4}x \Gamma^3 P_- q_-,
\label{d_sys3}\\
g_- = - \Gamma ^3(1-x^2 P_+)p_- + x\Gamma ^3 P_- q_+,
\label{d_sys4}\\
g_+ = - \frac{(1-x^2 P_+)^2}{1+x^2}\Gamma^2 (F_{\rm d} z),
\label{d_sys5}\\
g_- = -\frac{8 \lambda \sigma_{\rm M}(1-x^2 P_+)}{1+x^2}\frac{r_{\perp}^2}{r_{\rm jet}^2}
\left(\delta  - \frac{1}{2} \, \frac{\Omega_F}{\Omega_0} f\right),
\label{d_sys6}\\
\zeta = -\frac{A}{\sigma_{\rm M}} \Gamma^2 (F_{\rm d} z) + 4K \frac{x x_0}{1+x^2}\delta
+ 2K\frac{1-x^2 P_+}{1+x^2} f,
\label{d_sys7}\\
\frac{1}{r_{\perp}}\frac{\partial}{\partial r_{\perp}}\left[ r_{\perp}\frac{\partial}{\partial r_{\perp}}(r_{\perp}^2\delta)\right]+r_{\perp}^2\frac{\partial^2}{\partial z^2}\delta
-\frac{1}{r_{\perp}}\frac{\partial}{\partial r_{\perp}}(r_{\perp}^2 \zeta) =
\nonumber\\
= -2\lambda p_- + 4Kp_+,
\label{d_sys8}\\
-r_{\perp}^2 \frac{\partial^2}{\partial z^2}(f)
- r_{\perp}\frac{\partial}{\partial r_{\perp}}\left[\frac{1}{r_{\perp}}
\frac{\partial}{\partial r_{\perp}}(r_{\perp}^2 f) \right] =
\nonumber\\
= 4\lambda x x_0 p_- - 8x_0 K q_+,
\label{d_sys9}
\end{eqnarray}

As is shown in Appendix~\ref{A1}, the system of equations (\ref{d_sys1})--(\ref{d_sys9}) can be
rewritten as two second-order ordinary differential equations (\ref{x0first1})--(\ref{x0first2})
for $D = x_{0}^2 \delta$ and $F = x x_{0} f$ resulting in outside the light cylinder
\begin{eqnarray}
\frac{{\rm d^2}}{{\rm d} x_{0}^2} \left(D - \frac{F}{2}\right)
- \frac{16 \lambda^2 \sigma_{\rm M}}{\Gamma^3 x_{\rm jet}^2} \left(D - \frac{F}{2}\right)
+ \dots = 0.
\end{eqnarray}
Hence, the physical branch of equations (\ref{x0first1})--(\ref{x0first2}) corresponds to fast
diminishing solution $(D-F/2) \rightarrow 0 $ with the spacial scale
$\Delta r_{\perp} \ll r_{\rm jet}$, where
\begin{eqnarray}
\Delta r_{\perp} = \frac{\Gamma^{3/2}}{4 \lambda \sigma_{\rm M}^{1/2}} \, r_{\rm jet}.
\label{kala}
\end{eqnarray}

Thus, for $\Delta r_{\perp} \ll r_{\rm jet}$ (and for $\lambda \sigma_{\rm M} \gg 1$) one can
neglect l.h.s. of Eqn. (\ref{d_sys6}). As we see, in this case we return to one-fluid MHD
condition (\ref{c_one2}). Finding now $q_{+}$ from (\ref{d_sys3}) and  $\zeta$ from
(\ref{d_sys2}), we obtain two equations for $p_-$ and $\delta$
\begin{eqnarray}
2 \lambda p_{-}
- \frac{4Kx x_0 P_{+}}{(1-x^2 P_{+})} \,
\frac{1}{r_{\perp}}\frac{\partial}{\partial r_{\perp}}(r_{\perp}^2\delta)
\nonumber\\
+  \frac{16K^2(x^2+1-x^2 P_{+})x_0^2 P_{+}}{(1+x^2)(1-x^2 P_+)} \delta=
\nonumber\\
= \frac{4K x x_0 P_{+}}{(1 - x^2 P_{+})} \frac{A \Gamma^2}{\sigma_{\rm M}}  (F_{\rm d} z)
- 2\frac{A \Gamma^2}{\sigma}  (F_{\rm d} z),
\label{firsteq}\\
4\lambda x x_0 p_{-} -\frac{8 K x_0^2}{(1-x^2 P_{+})} \,
\frac{1}{r_{\perp}}\frac{\partial}{\partial r_{\perp}}(r_{\perp}^2\delta)
\nonumber\\
+ \frac{32 K^2 x_0^3 x}{(1-x^2 P_+)(1+x^2)}\delta =
 \frac{8 K x_{0} x}{(1-x^2 P_{+})}  \frac{A \Gamma^2}{\sigma} (F_{\rm d} z).
\label{secondeq}
\end{eqnarray}
Here
\begin{equation}
A(r_{\perp}) = \frac{r_{\rm jet}^2}{r_{\perp}^2}
\left[1-\frac{(1-x^2 P_{+})^2}{1+x^2} \right]\frac{\Omega_0}{\Omega_{\rm F}},
\label{A}
\end{equation}
so that $x^2 A \sim x_{\rm jet}^2 \gg 1$ ($A \sim 1$ for $x \sim x_{\rm jet}$),
and we neglect all the terms containing $\partial^2/\partial z^2$ (for small
$F_{\rm d}$ the derivatives along the jet are small), $x^{-2}$ and
$(1-x^2 P_{+}) \ll 1$. The full version is given in Appendix~\ref{A1}.

\subsubsection{Qualitative consideration}

At first, let us discuss the result obtained above qualitatively; the appropriate
numerical evaluations will be given in the next section. First, evaluating
$r_{\perp}^{-1}\partial (r_{\perp}^2\delta)/\partial r_{\perp}$  as $\delta$,
one can obtain
\begin{eqnarray}
\delta & = & k_{\delta} \, \frac{A}{\sigma_{\rm M}} \Gamma^2 (F_{\rm d}z),
\label{delta} \\
p_{-} & = &  \frac{k_{p}}{\lambda \sigma_{\rm M}} \, \frac{K A}{ (1-x^2 P_{+})}
\Gamma^2 (F_{\rm d}z),
\label{p-}
\end{eqnarray}
where $k_{\delta} \sim k_{p} \sim 1$. As we see, expression (\ref{delta}) for $\delta$
together with clear condition $|\delta| \sim 1$ for the full damping of the Poynting flux
reproduces immediately our evaluation (\ref{scale}) for the length
\mbox{$L_{\rm dr} = \sigma_{\rm M} m_{\rm e}c^2/F_{\rm drag}$;} now it can be rewritten
as
\begin{equation}
 L_{\rm dr} \sim \frac{\sigma_{\rm M}}{\Gamma^2 F_{\rm d}}.
\label{gminus}
\end{equation}

\begin{figure}
\includegraphics[scale=0.45]{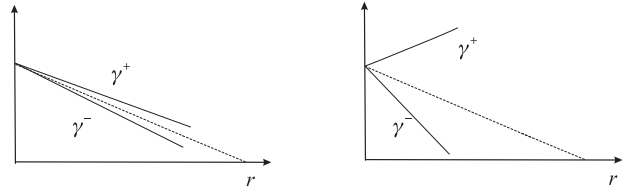}
\caption{
Hydrodynamical ($|g_{-}| \ll |g_{+}|$)  and non-hydro\-dy\-namical ($|g_{-}| > |g_{+}|$)
regimes of drag action. In the first case Lorentz-factors of electrons $\gamma^{-}$ and positrons
$\gamma^{+}$ actually coincide with the mean value $\Gamma$. In the last case one of the
species accelerates while another one decelerates more efficiently than by action of the
drag force only resulting in full stop at some point.
}
\label{fig02}
\end{figure}

Further, using expression (\ref{p-}) for $p_{-}$ together with (\ref{d_sys1}) and
(\ref{d_sys3}) one can obtain
\begin{equation}
g_{-} \sim \frac{A}{\lambda \sigma_{\rm M}} \, \Gamma^5 \, (F_{\rm d}z).
\label{gminusbis}
\end{equation}
Together with (\ref{d_sys5}) it gives
\begin{equation}
\frac{g_{-}}{g_{+}} \sim \frac{1}{\lambda \sigma_{\rm M}} \,
\frac{(1 + x^2)A}{(1 - x^2P_{+})^2} \, \Gamma^3.
\label{answer}
\end{equation}
Relation (\ref{answer}) is actually our main result separating hydrodynamical and
non-hydrodynamical regime of the drag force action. Indeed, for large enough multiplicity
$\lambda > \lambda_{\star}$, where
\begin{equation}
\lambda_{\star} = \frac{x_{\rm jet}^2 \Gamma^3}{\sigma_{\rm M}(1-x^2P_{+})^2}.
\label{lama}
\end{equation}
the difference in Lorentz-factors of electrons and positrons is negligible, and we deal with
one-fluid MHD flow. On the other hand, for $|g_{-}| > |g_{+}|$ the drag force $F_{\rm drag}$
is smaller than the electrostatic one $eE_{\parallel}$. As a result, as is shown on Fig.~\ref{fig02},
one of the species accelerates while another one decelerates more efficiently than by the action
of the drag force only resulting in full stop at some point. It is clear that in the last case
the very hydrodynamical description it now impossible. As the condition (\ref{lama}) can be
rewritten as
\begin{equation}
\lambda \sigma_{\rm M} =
\frac{x_{\rm jet}^2 \Gamma^5}{\Gamma_{0}^2},
\label{crit}
\end{equation}
we see that according to (\ref{newsigma}) non-hydrodynamical regime can be realised for small
$W_{\rm tot} < W_{\star}$, where
\begin{equation}
W_{\star} = \frac{x_{\rm jet}^4 \Gamma^{10}}{\Gamma_{0}^{4}} W_{\rm A},
\label{Wstar}
\end{equation}
where again $W_{\rm A} = m_{\rm e}^{2}c^{5}/e^{2} \approx 10^{17}$ erg/s.
The corresponding  Poyting flux is less than
\begin{equation}
S_{\star} = \frac{x_{\rm jet}^2 \Gamma^{10}}{\Gamma_{0}^{4}} W_{\rm A},
\label{Wstarbis}
\end{equation}
Accordingly, in the non-hydrodynamical regime the distance $L_{\rm st}$ to the stop
point can be evaluated as
\begin{equation}
L_{\rm st} \sim \frac{\lambda \sigma_{\rm M}}{\Gamma^4 F_{\rm d}}.
\label{lz}
\end{equation}

Finally, in one-fluid approximation corresponding to condition $|g_{-}| \ll |g_{+}|$
we can write down
\begin{equation}
\frac{\partial}{\partial z}\Gamma =
- \frac{(1-x^2 P_+)^2}{1+x^2} F_{\rm d}\Gamma^2.
\label{mmm}
\end{equation}
Certainly, it is possible to use this solution for small disturbance of the Lorentz-factor
$\Gamma$ only. Nevertheless, we again can evaluate the distance $L_{\Gamma}$ of the essential
diminishing of the bulk particle energy $m_{\rm e}c^2 \Gamma$ of the  motion on the scale $L$
\begin{equation}
L_{\Gamma} \sim \frac{x_{\rm jet}^2 \Gamma}{\Gamma_{0}^2 F_{\rm d}}. \,
\label{gamev}
\end{equation}
As we see, this distance is much larger that $L_{\rm dr}$. It is not surprising as
in the linear approximation, as was already stressed, the particle energies remain
actually constant. For this reason it is impossible to use the value $L_{\Gamma}$
as the evaluation of the retardation length.

\subsubsection{Quantitative consideration}

Finally, below we present the result of numerical integration of the linear system
(\ref{firsteq})--(\ref{secondeq}). Neglecting $p_{-}$ and the derivatives
$\partial/\partial z$, one can obtain the second-order ordinary differential equation
for determination $\delta$ only. It looks like (see Appendix~\ref{A1} for more detail)
\begin{eqnarray}
2x\frac{{\rm d}}{{\rm d} x_0}\left[x_0 \frac{{\rm d}}{{\rm d} x_0}D \right]-
2x_0\frac{{\rm d}}{{\rm d} x_0} \left[ \frac{1}{x_0}\frac{{\rm d}}{{\rm d} x_0 }
\left(\frac{\Omega_0}{\Omega_{\rm F}}D\right) \right]+
\nonumber\\
8x\frac{{\rm d}}{{\rm d} x_0}
\left[K\frac{(x_0x +\Omega_{0}/\Omega_{\rm F}-x^2P_{+}\Omega_{0}/\Omega_{\rm F})}{(1+x^2)}D \right]+
\nonumber\\
8K x_0\frac{{\rm d}}{{\rm d} x_0}D  -\frac{32K^2 x_0 (x^2+1-x^2 P_+)}{x(1+x^2)}D
\nonumber\\
= - 2x\frac{{\rm d}}{{\rm d} x_0 } \left[x_0^2 {\cal G}\right]-8Kx_0^2{\cal G},
\label{main_delta_B}
\end{eqnarray}
where $D=x_0^2 \delta$, ${\cal G}= A \Gamma^2 (F_{\rm d}z)/\sigma_{\rm M}$ and again
$x_{0} = \Omega_{0} r_{\perp}/c$ and $x = \Omega(r_{\perp}) r_{\perp}/c$. As to angular
velocity profile $\Omega_{\rm F}(r_{\perp})$ which determines the coefficient $K$
(\ref{c_K}), we use the simplest relation
\begin{equation}
\Omega_{\rm F}(r_{\perp}) = \Omega_{0}\left(1 - \frac{r_{\perp}^2}{r_{\rm jet}^2}\right)
\label{OOmeg}
\end{equation}
corresponding to zero total electric charge and electric current within the jet
$\Omega_{\rm F}(r_{\rm jet}) = 0$.

Additional remarks are to be done to boundary conditions. As is shown in Appendix~\ref{A2},
to avoid longitudinal electric field on the jet axis it is necessary to put $D(0) = 0$.
Together with the regularity condition at the light cylinder $x = 1$ it helps us to obtain
the full solution of a problem.

\begin{figure}
\includegraphics[scale=1.3]{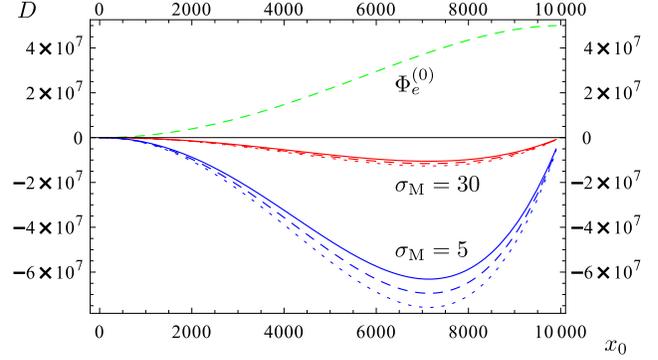}
\caption{
Solution $D = x_{0}^2 \delta$  of Eqn. (\ref{main_delta_B}) for $x_{\rm jet} = 10^4$ and for
different values $\sigma_{\rm M}$. Solid, dashed and dotted lines of the same color correspond to three
different value of $F_{\bf d} z$: 1, 1.1, 1.2 respectively. Upper curve corresponds to undisturbed electric
potential $\Phi_{\rm e}^{(0)}$.
}
\label{fig03}
\end{figure}

\begin{figure}
\includegraphics[scale=1.1]{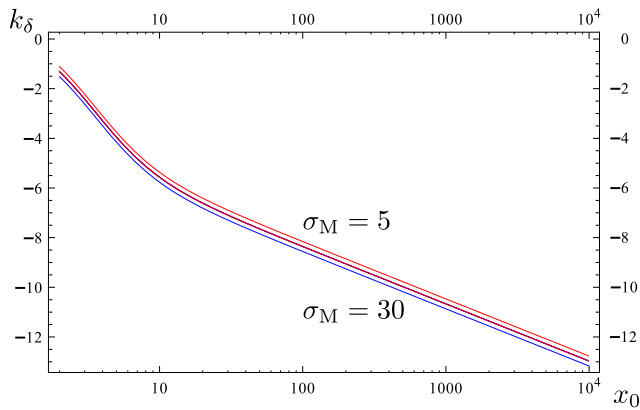}
\caption{Dimensionless function $k_{\delta}$ which actually does not depend on magnetization
parameter $\sigma_{\rm M}$.
}
\label{fig04}
\end{figure}

As it is shown in Fig.\ref{fig03}, solution of equation (\ref{main_delta_B}) gives negative
values for the disturbance of the electric potential $\delta$. This just implies that the
disturbance $\delta$ resulting from drag force compensates gradually the electric potential of
the jet (upper curve).  Moreover, as is shown on Fig.~\ref{fig04}, our evaluation (\ref{delta})
reproduses good enough the exact solution of Eqn.~(\ref{main_delta_B}).

Finally, as, according to (\ref{c_one2}), disturbance of magnetic surfaces $f$ is to be negative
as well, one can rewrite the magnetic flux $\Psi(r_{\perp}, z)$ (\ref{c_cor3}) as
\begin{equation}
\Psi (r_{\perp}, z)=\pi B_0 r_{\perp}^2 \left( 1-Cz \right),
\label{rewritten}
\end{equation}
where $C > 0$. It leads to appearance of a positive radial component of the magnetic field
$B_{r}$ (\ref{c_cor5}), i.e., to decollimation of the jet{\footnote{We consider here
the case $B_{z} > 0$.}}. But as one can easily check, the width of the jet increases
essentially only for $\delta \sim 1$ when almost all electromagnetic energy will be
transferred into IC photons.

\section{Astrophysical applications and discussions}
\label{sect:diss}

Thus, we have demonstrated how for simple geometry it is possible to determine the small
correction of the one-fluid ideal outflow resulting from radiation drag force. In comparison
with the paper by Li et al (1992), both the disturbance of magnetic surfaces and electric potential
were included into consideration self-consistently. As a result, the possibility arises to find
the tendency of the drag action on the ideal MHD magnetically dominated outflow as well as to
evaluate the conditions when this disturbance becomes large.

Let us try now to evaluate the real role of radiation drag in dynamics of relativistic jets in
active galactic nuclei. As the energy density $U_{\rm iso}$ at the distance $R$ from the
'central engine' with the total luminosity $L_{\rm tot}$ can be estimated as
$U_{\rm tot} \sim 10^{-3}$ erg/cm$^3$ at the distance $R = 10$ pc. Assuming that
$U_{\rm iso} \sim 0.1 U_{\rm tot} \sim 10^{-4}$ erg/cm$^3$, see, e.g.,~\citet{Joshi}),
one can evaluate the length of hydrodynamical retardation $L_{\rm dr}$ given by (\ref{scale}) as
\begin{equation}
L_{\rm dr}  \sim 300
\left(\frac{\sigma_{\rm M}}{10}\right)
\left(\frac{\Gamma}{10}\right)^{-2}
\left( \frac{L_{\rm tot}}{10^{-4} \, {\rm erg/cm^{3}}}\right)^{-1}
{\rm pc}.
\end{equation}
Thus, for $\Gamma \sim \sigma_{\rm M} \sim 10$ obtained recently by~\citet{NBKZ} from analysis
of about 100 sources using core-shift technics, the distance is quite reasonable to explain the
observable retardation on the scale $R \sim 100$ pc.

On the other hand, on the scale $R \sim 10$ kpc, corresponding to dimension of the galaxy,
where $U_{\rm iso} \sim 10^{-10}$ erg/cm$^3$, the retardation length $L_{\rm dr}$ is too
large to prevent the jet material reaching the lobes. To conclude, in our opinion, isotropic
photon field can be considered as one of the possible reason of jet deceleration in active
galactic nuclei.

Finally, it is very interesting to discuss the photon drag action in connection with~\citet{FR}
classification. At first glance, deceleration is to be more effective in FRII objects, i.e., in
objects in which the ambient radiation field is more intense. But as was demonstrated above, in
objects with higher magnetization $\sigma_{\rm M}$, the drag force acts indirectly diminishing
mainly the electromagnetic flux. As far as FRI sources, in which one can expect particle dominated
flow in parsec scales, drag is to be much more effective. We are going to consider the statistics
of the sources in Paper III.

\section*{Acknowledgments}

We would like to acknowledge M.~Barkov, E.~Derishev, Ya.~Istomin  and especially N.~Zakamska
for useful comments. We also thank the anonymous referee for his/her helpful remarks. This
work was supported by Russian Science Foundation, grant 16-12-10051.

\appendix

\section{Linearization in the drift approximation}
\label{A1}

In this Appendix we determine the linear disturbances to the cylindrical
drag-free flow in the drift approximation. First, using the definitions
(\ref{c_cor5})--(\ref{c_cor7}) for the total magnetic field ${\bf B}$ and clear
expressions \mbox{${\bf V}_{\parallel} = ({\bf V}{\bf B}){\bf B}/{\bf B}^2$}
and \mbox{${\bf V}_{\perp}={\bf V}-{\bf V}_{||}$} for any vector ${\bf V}$,
we obtain for the perpendicular components of vectors ${\bf e} = {\bf E}/B_{0}$
and ${\bf F}_{\rm dr}$
\begin{eqnarray}
e_{\perp}^r & = & -x-x_0 \frac{1}{r_{\perp}} \frac{\partial}{\partial r_{\perp}}
\left( r_{\perp} ^2 \delta \right),
\label{eperpr} \\
e_{\perp }^{\varphi} & = & \frac{x}{(x^2+1)}\left(\frac{1}{2}x r_{\perp}
\frac{\partial f}{\partial z} -x_0 r_{\perp} \frac{\partial \delta}{\partial z}\right),
\label{eperpf} \\
e_{\perp}^z & = & -x_0 r_{\perp}\frac{\partial \delta}{\partial z}-\frac{1}{x^2+1}
\left(\frac{1}{2}x r_{\perp}\frac{\partial f}{\partial z}- x_0 r_{\perp}
\frac{\partial \delta}{\partial z} \right),
\label{eperpz}\\
F_{\perp }^{\varphi} & = & -\frac{F_{\rm d}\gamma^2}{\sqrt{1-2\xi_z+\xi_{\varphi}^2}}
\frac{(x+\xi_{\varphi})}{1+x^2},
\label{fperpf} \\
F_{\perp }^z & = & -\frac{F_{\rm d}\gamma^2 x}{\sqrt{1-2\xi_z+\xi_{\varphi}^2}}
\frac{(x+\xi_{\varphi})}{1+x^2},
\label{fperpz} \\
F_{\perp }^r & = & 0.
\end{eqnarray}
Using now these expressions we can obtain for $r$-component of the drift velocity
\begin{equation}
\xi_r^{\rm dr} = - \frac{(x+\xi_{\varphi})}{(x^2+1)} 
F_{\rm d} \gamma^2 -
\frac{xx_0 r_{\perp}}{(x^2+1)} \frac{\partial \delta}{\partial z},
\end{equation}
and for $r$-component of longitudinal velocity
\begin{equation}
(\xi_{\parallel})_r = -\frac{1}{2} \, \frac{(1-x\xi_{\varphi})}{(1+x^2)}
r_{\perp}\frac{\partial f}{\partial z}.
\end{equation}
Substituting these expressions into (\ref{c_sys7})--(\ref{c_sys8}) and
remembering that $\xi_{\varphi} \approx x P_+$, we result in (\ref{gamma}).

Further, combining (\ref{c_sys3}), (\ref{c_sys7})--(\ref{c_sys8}), and (\ref{gamma}),
one can find
\begin{eqnarray}
r_{\perp}\frac{\partial \zeta}{\partial z} =4K r_{\perp} \frac{\Omega_0}{\Omega_F}
\, \frac{\partial \delta}{\partial z}
\nonumber\\
-F_{\rm d} \Gamma^2\left[1-\frac{(1-x^2 P_{+})^2}{(1+x^2)}\right]
\frac{\Omega_0}{\Omega_F} \, \frac{r_{\rm jet}^2}{r_{\perp}}+
\nonumber\\
+\frac{4K \Omega_0 }{r_{\perp} \Omega_F} \, \frac{(1-x^2 P_{+})}{(1+x^2)}
\left(-r_{\perp}^2 \frac{\partial \delta}{\partial z}
+r_{\perp}^2\frac{\Omega_F}{2 \Omega_0}
\frac{\partial f}{\partial z}\right),
\end{eqnarray}
where we put $(\gamma^+)^2+(\gamma^-)^2 = 2 \Gamma^2$. Integrating it, we obtain
\begin{eqnarray}
\zeta = \frac{A}{\sigma}\int F_{\rm d}\Gamma^2 {\rm d}z
+4K \frac{x x_0}{(x^2+1)}\delta
+ 2K \frac{(1-x^2 P_{+})}{(1+x^2)} f,
\end{eqnarray}
where $A$ is given by (\ref{A}). Finally subtracting equation (\ref{c_sys6}) from
(\ref{c_sys5}) and neglecting l.h.s., one can  obtain the following expression
\begin{equation}
q_{+} = x p_{+} + \frac{1}{R_{\rm L}} \frac{\partial }{\partial r_{\perp}} (r_{\perp}^2 \delta)
- x_0 \zeta,
\label{q+}
\end{equation}
where again $R_{\rm L} = c/\Omega_{0}$ is the light cylinder radius.

Finally, using definitions (\ref{def1})--(\ref{def3}) and expressing $\gamma^{+}$
and $\gamma^{-}$ through $\Gamma$ and $G$, we have
\begin{eqnarray}
\frac{1}{(\Gamma + G/2)^2} = 2\left(P_{+} + \frac{P_{-}}{2}\right) - \left(Q_{+}
+ \frac{Q_{-}}{2}\right)^2,
\label{sum}\\
\frac{1}{(\Gamma - G/2)^2} = 2\left(P_{+} - \frac{P_{-}}{2}\right) - \left(Q_{+}
- \frac{Q_{-}}{2}\right)^2.
\label{dif}
\end{eqnarray}
They give for $G \ll \Gamma$
\begin{equation}
G = - \Gamma ^3 (1-x^2 P_+)P_{-},
\label{gamma--}
\end{equation}
and
\begin{equation}
g_{-} = -(1-x^2 P_+)\Gamma ^3 p_{-} + x P_- \Gamma ^3 q_+.
\end{equation}
These relations lead to a system of equations (\ref{d_sys1})--(\ref{d_sys9}).

As a result, expressing $p_+$ from (\ref{d_sys3}) and substituting it together
with $\zeta$ in (\ref{d_sys2}), we obtain
\begin{eqnarray}
q_+ = -\frac{x}{\Gamma ^3(1-x^2 P_+)}g_{+}
+ \frac{1}{(1-x^2 P_{+})}\frac{1}{R_{\rm L}}\frac{\partial}{\partial r_{\perp}}(r_{\perp}^2 \delta)
\nonumber\\
+ \frac{x_0}{(1-x^2 P_{+})}\left[\frac{A}{\sigma}\Gamma^2 (F_{\rm d} z)
- \frac{4K x x_0}{(1+x^2)}\delta - \frac{2K(1-x^2 P_{+})}{(1+x^2)} f \right].
\end{eqnarray}
Put it in (\ref{d_sys9}), we obtain
\begin{eqnarray}
4 \lambda x x_0 p_{-} + r_{\perp}\frac{\partial }{\partial r_{\perp}}
\left[\frac{1}{r_{\perp}}\frac{\partial}{\partial r_{\perp}}(
r_{\perp}^2 f)\right]+\frac{16 K^2 x_{0}^2}{(1+x^2)} f +
\nonumber\\
\frac{32K^2 x_0^3 x}{(1-x^2 P_+)(1+x^2)}\delta-
\frac{8Kx_{0} ^2}{(1-x^2 P_{+})}
\frac{1}{r_{\perp}}\frac{\partial}{\partial r_{\perp}}(r_{\perp}^2 \delta) +
\nonumber\\
r_{\perp}^2 \frac{\partial^2 f }{\partial z^2}
=  \frac{8K x_0^2}{(1-x^2P_{+})}\frac{A}{\sigma}{ \cal G} -
\frac{8 K x x_0(1-x^2 P_{+})}{1+x^2}\frac{\cal G}{\Gamma^3},
\label{first}
\end{eqnarray}
where ${\cal G} = \Gamma^2 (F_{\rm d}z)$. Besides,  substituting (\ref{d_sys7}) into
(\ref{d_sys8}) and expressing $p_+$, we get the second equation
\begin{eqnarray}
2 \lambda p_{-} + \frac{16 K^2 x_0^2 P_{+}(x^2+ 1 -x^2 P_{+})}{(1+x^2)(1-x^2 P_+)}
\delta+ r_{\perp}^2\frac{\partial^2 \delta}{\partial z^2}
\nonumber\\
+\frac{1}{r_{\perp}}\frac{\partial}{\partial r_{\perp}}
\left[r_{\perp}\frac{\partial}{\partial r_{\perp}} (r_{\perp}^2\delta) \right]
- \frac{4}{r_{\perp}}\frac{\partial}{\partial r_{\perp}}
\left[r_{\perp}^2 K \frac{x x_0}{(1+x^2)}\delta \right]
\nonumber\\
-\frac{2}{r_{\perp}} \frac{\partial}{\partial r_{\perp}}
\left[ r_{\perp}^2 K \frac{(1-x^2 P_{+})}{1+x^2} f \right]
\nonumber\\
-\frac{4 K x x_0 P_+}{(1-x^2 P_{+})} \frac{1}{r_{\perp}}
\frac{\partial}{\partial r_{\perp}}(r_{\perp}^2 \delta)
 = - \frac{1}{r_{\perp}}\frac{\partial}{\partial r_{\perp}}
(r_{\perp}^2 \frac{A}{\sigma} {\cal G})
 \nonumber\\
+ \frac{4K x x_0 P_+}{(1-x^2 P_{+})} \frac{A}{\sigma}{\cal G}
+ \frac{4K (1-x^2 P_{+})}{\Gamma^3(1+x^2)} {\cal G}.
\label{second}
\end{eqnarray}
Neglecting now longitudinal derivatives $\partial^2/\partial z^2$ one can rewrite the
system of equations (\ref{first})--(\ref{second}) as two second-order ordinary differential
equations for $D = x_{0}^2 \delta$ and $F = xx_{0} f$
\begin{eqnarray}
\frac{{\rm d^2}D}{{\rm d} x_{0}^2}  =  -\frac{1}{x_{0}} \, \frac{{\rm d}D}{{\rm d} x_{0}}
+\frac{1}{x_{0}} \, \frac{{\rm d}Y}{{\rm d} x_{0}} - 2\lambda p_{-} + 4 K p_{+},
\label{x0first1} \\
\frac{{\rm d^2}F}{{\rm d}x_{0}^2}=-\left[ \frac{1}{x_0}+2x\frac{{\rm d}}{{\rm d}x_{0}}
\left(\frac{1}{x}\right)\right]\frac{{\rm d}F}{{\rm d} x_{0}} -4\lambda x^2 p_{-}
\nonumber\\
+8Kxq_{+}-\left[x\frac{{\rm d^2}}{{\rm d}x_{0}^2}\left(\frac{1}{x}\right)+\frac{x}{x_0}
\frac{{\rm d}}{{\rm d} x_{0}}\left(\frac{1}{x}\right)-\frac{1}{x_{0}^2}\right]F.
\label{x0first2}
\end{eqnarray}
Here $Y = x_{0}^2 \zeta$,
\begin{eqnarray}
Y =  \frac{4Kxx_{0}}{(1+x^2)} D + \frac{2Kx_{0}(1-x^2P_{+})}{x(1+x^2)} F
-\frac{A x_{0}^2}{\sigma_{\rm M}} \, {\cal G},
\label{x0last1}
\end{eqnarray}
${\cal G} = \Gamma^2 (F_{\rm d}z)$, and
\begin{eqnarray}
\lambda p_{-}  =  \frac{8 \lambda^2 \sigma_{\rm M}}{\Gamma^3 x_{jet}^2 (1+x^2)}
\left(D-\frac{1}{2}F\right)+\frac{2KP_{+}x}{(1+x^2)}\frac{\cal G}{\Gamma^3}+
\nonumber\\
\frac{2KP_{+}}{(1-x^2P_{+})^2}\frac{\partial}{\partial x_{0}}D -
\frac{2KP_{+}}{x_0 (1-x^2P_{+})^2}Y,
\nonumber\\
p_{+}  =  \frac{(1-x^2P_{+})}{(1+x^2)}\frac{\cal G}{\Gamma^3}+\frac{xP_{+}}{(1-x^2P_{+})}
\frac{\partial}{\partial x_0}D - \frac{xP_{+} Y}{x_0 (1-x^2P_{+})},
\nonumber\\
q_ {+} = \frac{(1-x^2 P_{+})x}{\Gamma^3 (1+x^2)} {\cal G}
+\frac{1}{(1-x^2 P_{+})} \frac{{\rm d}D}{{\rm d} x_{0}} -\frac{Y}{x_{0}(1-x^2 P_{+})}.
\label{x0last2}
\end{eqnarray}
Outside the light cylinder $x_{0} \gg 1$  it gives
\begin{eqnarray}
\frac{{\rm d^2}}{{\rm d} x^2} \left(D - \frac{F}{2}\right)
- \frac{16 \lambda^2 \sigma_{\rm M}}{\Gamma^3 x_{\rm jet}^2} \left(D - \frac{F}{2}\right)
+ \dots = 0.
\end{eqnarray}
Hence, the physical branch of equations (\ref{x0first1})--(\ref{x0first2}) corresponds to fastly
diminishing solution $(D-F/2) \rightarrow 0 $ with the spacial scale
$\Delta x \ll 1$
\begin{eqnarray}
(\Delta x)^2 = \frac{\Gamma^3 x_{\rm jet}^2}{16 \lambda^2 \sigma_{\rm M}}.
\label{Akala}
\end{eqnarray}
Finally, for $D = F/2$, i.e., in the one-fluid MHD approximation ($E_{\parallel} = 0$)
Eqns. (\ref{first})--(\ref{second}) result in (\ref{firsteq})--(\ref{secondeq}).
Expressing now $p_{-}$ from (\ref{first}) and put it into (\ref{second}), we finally obtain
Eqn. (\ref{main_delta_B}).

\section{Toy model}
\label{A2}

Boundary condition $D(0) = 0$ to equation (\ref{main_delta_B}) is actually one of the
key nontrivial property of the solution discussed above. The point is that any finite
central engine with dipole-like magnetic field procuces quadrupole electric field so
that the potential difference between its magnetic pole and infinity does not vanish. On
the contrary, our solution corresponds to zero (more exactly, very small) electric
field $E_{z}$ along the rotational axis.

To demonstrate the very possibility for the longitudinal electric field $E_{z}$ to be
small, let us write down by hand the electric potential in the region $z > 0$ (and
vanishing at infinity) in the form
\begin{equation}
\Phi_{\rm e} = \frac{\Omega_{0}B_{0}}{c} \, r_{\rm jet}^2
\left({\cal C} +\frac{1}{2} \, \frac{r_{\perp}^2}{r_{\rm jet}^2}
- \frac{1}{4} \, \frac{r_{\perp}^4}{r_{\rm jet}^4}\right)
\exp\left(-\frac{z^2}{L_{\rm dr}^2}\right),
\end{equation}
where we use the expression (\ref{OOmeg}) for angular velocity $\Omega_{\rm F}$. For
$L_{\rm dr} \rightarrow \infty$ it corresponds to electric field $E_{r}^{(0)}$ (\ref{E&B})
for arbitrary constant ${\cal C}$. In particular, it gives the same zero-order charge
density $\rho_{\rm e}$ (\ref{rhoe}). As was already stressed, ${\cal C} <0$ ($|{\cal C}| \sim 1$)
for $L_{\rm dr} \rightarrow 0$, i.e. for spatially limited quadrupole charge distribution.

On the other hand, for finite $L_{\rm dr}$ the disturbance of charge density in the
vicinity of the rotational axis for $z \sim r_{\rm jet}$ depends drastically on constant
${\cal C}$. Indeed, the additional charge density can be divided into two terms, namely,
the negative part
\begin{equation}
\delta \rho_{\rm e}^{(1)} = - |{\cal C}|\frac{\Omega_{0}B_{0}}{2 \pi c} \,
\frac{r_{\rm jet}^2}{L_{\rm dr}^2}
\end{equation}
existing for ${\cal C} \neq 0$ (and producing electric field $E_{z} < 0$ along the
rotation axis opposite the particle flow), and the positive one
\begin{equation}
\delta \rho_{\rm e}^{(2)} = \frac{\Omega_{0}B_{0}}{2 \pi c} \, \frac{z^2}{L_{\rm dr}^2}
\end{equation}
having the same order of magnitude on the scale $z \sim r_{\rm jet}$. This implies
that the small redistribution of the charge density in the base of the flow indeed can
screen the longitudinal electric field along the jet.


\begin{thebibliography}{99}
\bibitem[\protect\citeauthoryear{Aharonian, Bogovalov \& Khangulian}{2012}]{nature}
Aharonian~F.A., Bogovalov~S.V, Khangulian~D., 2012, Nature, 482, 507
\bibitem[\protect\citeauthoryear{Appl \& Camenzind}{1992}]{A&C}
Appl~S., Camenzind~M., 1992, A\&A, 256, 354
\bibitem[\protect\citeauthoryear{Barkov \& Kommisarov}{2016}]{BarKom}
Barkov~M.V., Kommisarov~S.S., 2016, MNRAS, 458, 1939
\bibitem[\protect\citeauthoryear{Beal, Guillori \& Rose}{2010}]{MmSAI}
Beal~J,H., Guillori~J, Rose~D.V., 2010, Mem. Soc. Astron. Ital., 81, 404
\bibitem[\protect\citeauthoryear{Begelman, Blandford \& Rees}{1984}]{BBR84}
Begelman~M.C., Blandford~R.D., Rees~M.J., 1984, Rev. Mod. Phys., 56, 255
\bibitem[\protect\citeauthoryear{Benford}{1981}]{benford}
Benford~G., 1981,ApJ., 247, 792
\bibitem[\protect\citeauthoryear{Beskin}{2009}]{Beskinbook}
Beskin~V.S., 2009, MHD Flows in Compact Astrophysical Objects. Springer, Berlin
\bibitem[\protect\citeauthoryear{Beskin}{2010}]{Beskin-10}
Beskin~V.S., 2010, Physics-Uspekhi, 53, 1199
\bibitem[\protect\citeauthoryear{Beskin, Gurevich \& Istomin}{1993}]{BGI93}
Beskin~V.S., Gureich~A.V., Istomin~Ya.N., 1993, Physics of the Pulsar Magnetosphere.
Cambridge University Press, Cambridge
\bibitem[\protect\citeauthoryear{Beskin, Kuznetsova \& Rafikov}{1998}]{BKR}
Beskin~V.S., Kuznetsova~I.V., Rafikov~R.R., 1998, MNRAS, 299, 341
\bibitem[\protect\citeauthoryear{Beskin \& Nokhrina}{2006}]{BN-06}
Beskin~V.S., Nokhrina~E.E., 2006, MNRAS, 367, 375
\bibitem[\protect\citeauthoryear{Beskin \& Nokhrina}{2016}]{paper2}
Beskin~V.S., Nokhrina~E.E., 2016, MNRAS (in preparation)
\bibitem[\protect\citeauthoryear{Beskin \& Rafikov}{2000}]{BR2000}
Beskin~V.S., Rafikov~R.R., 2000, MNRAS, 313, 433
\bibitem[\protect\citeauthoryear{Beskin, Zakamska \& Sol}{2004}]{BZS-04}
Beskin~V.S., Zakamska~N.L., Sol~H., 2004, MNRAS, 347, 587
\bibitem[\protect\citeauthoryear{Bing \& Huirong}{2011}]{rec5}
Bing~Z., Huirong~Ya., 2011, ApJ, 726, 90Z
\bibitem[\protect\citeauthoryear{Blumenthal \& Gould}{1970}]{Blumenthal&Gould}
Blumethal~G.R., Gould~R.G., 1970, Rev. Mod. Phys., 42, 237
\bibitem[\protect\citeauthoryear{Bogovalov et~al.}{2008}]{BHeid1}
Bogovalov~S.V., Khangulyan~D.V., Koldoba~A.V., Ustyugova~G.V., Aharonian~F.A.,
2008, MNRAS, 387, 63
\bibitem[\protect\citeauthoryear{Bogovalov et~al.}{2012}]{BHeid2}
Bogovalov~S.V., Khangulyan~D.V., Koldoba~A.V., Ustyugova~G.V., Aharonian~F.A.,
2008, MNRAS, 419,
\bibitem[\protect\citeauthoryear{Bogovalov \& Tsinganos}{1999}]{BTs99}
Bogovalov~S.V., Tsinganos~K.,1999, MNRAS, 305, 211
\bibitem[\protect\citeauthoryear{Bucciantini et al}{2009}]{Bucc}
Bucciantini~N., Quataert~E., Metzger~B.D., Thompson~T.A., Arons~J., del Zanna~L.,
2006, ApJ, 396, 2038
\bibitem[\protect\citeauthoryear{Cerutti et~al.}{2015}]{Cerutti}
Cerutti~B., Philippov~A.A.., Parfrey~K., Spitkovsky~A., 2015, MNRAS, 448, 606
\bibitem[\protect\citeauthoryear{de la Cita et~al.}{2016}]{DelaCita}
de la Cita~V.M., Bosch-Ramon~V., Paredes-Fortuni~X., Khangulyan~D., Perucho~M. \ 2016,
A\&A (in press) ArXiv:1604.02070v1
\bibitem[\protect\citeauthoryear{Clausen-Brown et~al.}{2013}]{AGN3}
Clausen-Brown~E., Savolainen~T., Pushkarev~A.B., Kovalev~Y.Y., Zensus~J.A.\ 2013,
A\&A, 558, A144
\bibitem[\protect\citeauthoryear{Cohen et~al.}{2007}]{AGN2}
Cohen~M.H. et al, 2007, ApJ, 658, 232
\bibitem[\protect\citeauthoryear{Derishev et~al.}{2003}]{DAKK03}
Derishev~E.V., Aharonian~F.A., Kocharovsky~V.V., Kocharovsky~Vl.V., 2003,
Phys.~Rev.~D, 68, 043003
\bibitem[\protect\citeauthoryear{Double et~al.}{2004}]{Double}
Double~G.P., Baring~M.G., Jones~F.C., Ellison~D.C., 2004, ApJ, 600, 485
\bibitem[\protect\citeauthoryear{Drenkhahn \& Spruit}{2002}]{rec2}
Drenkhahn~G., Spruit~H.C., 2002, A\&A, 391, 1141
\bibitem[\protect\citeauthoryear{Fanaroff \& Riley }{1974}]{FR}
Fanaroff~B.L., Riley~J.M., 1974, MNRAS, 167, 31P
\bibitem[\protect\citeauthoryear{Ferraro}{1937}]{Ferraro}
Ferraro~V.C.A., 1937, MNRAS, 97, 458
\bibitem[\protect\citeauthoryear{Gabuzda, Murrey \& Cronin}{2005}]{Gab92}
Gabuzda~D., Murrey~E., Cronin~P. 2005, MNRAS, 351, 8
\bibitem[\protect\citeauthoryear{Golan \& Levinson}{2015}]{rec4}
Golan~O., Levinson~A., 2015, ApJ, 809, 23
\bibitem[\protect\citeauthoryear{Goldreich \& Julian}{1969}]{GJ1}
Goldreich~P., Julian~W.H. 1969, ApJ, 157, 869
\bibitem[\protect\citeauthoryear{Goldreich \& Julian}{1970}]{GJ2}
Goldreich~P., Julian~W.H. 1969, ApJ, 160, 971
\bibitem[\protect\citeauthoryear{Hardee \& Norman}{1988}]{hardee2}
Hardee~P.E., Norman~M.L., 1988, ApJ, 334, 70
\bibitem[\protect\citeauthoryear{Heyvaerts \& Norman}{1989}]{HN}
Heyvaerts~J., Norman~J., 1989, ApJ, 347, 1055
\bibitem[\protect\citeauthoryear{Hirotani \& Okamoto}{1998}]{Hir98}
Hirotani~K., Okamoto~I., 1998, ApJ, 497, 563
\bibitem[\protect\citeauthoryear{Homan et~al.}{2015}]{MOJAVE12}
Homan~D.C., Lister~M.L., Kovalev~Y.Y., Pushkarev~A.B.,
Savolainen~T, Kellermann~K.I., Richards~J.L., Ros~E., 2014, ApJ, 798, 16
\bibitem[\protect\citeauthoryear{Istomin \& Pariev}{1994}]{IP94}
Istomin~Ya.N., Pariev~V.I., 1994, MNRAS, 267, 629
\bibitem[\protect\citeauthoryear{Joshi et al}{2014}]{Joshi}
Joshi~M., Marscher~A.P., Bottcher~M.,  2014, ApJ, 785, 132
\bibitem[\protect\citeauthoryear{Kardashev et~al.}{2014}]{AGN4}
Kardashev~N.S., Novikov~I.D, Lukash~V.N. et al. 2014, Phys. Uspehkhi, 57, 1199
\bibitem[\protect\citeauthoryear{Kennel, Fujimura, Okamoto}{1976}]{KFO76}
Kennel~C.~F., Fujimura~F.~S., Okamoto~I., 1983, Geophys. Astrophys.
Fluid Dyn., 26, 147
\bibitem[\protect\citeauthoryear{Komissarov}{1994}]{Komissarov-94}
Komissarov~S., 1994, MNRAS, 269, 394
\bibitem[\protect\citeauthoryear{Komissarov et~al.}{2007}]{num1}
Komissarov~S., Barkov~M., Vlahakis~N., K\"onigl~A., 2007, MNRAS, 380, 51
\bibitem[\protect\citeauthoryear{Levinson \& Globus}{2016}]{LevGlo16}
Levinson~A. Globus~N. 2016, MNRAS, 458, 2269
\bibitem[\protect\citeauthoryear{Lery et al}{1998}]{Lery98}
Lery~T., Heyvaerts~J., Appl~S., Norman~C.A., 1998, A$\&$A, 337, 603
\bibitem[\protect\citeauthoryear{Li, Begelman \& Chiueh}{1992}]{LBC92}
Li~Z.-Y., Begelman~M., Chiueh~T., 1992, ApJ, 384, 567
\bibitem[\protect\citeauthoryear{Lobanov}{1998}]{AGN1}
Lobanov~A.P., 1998, A\&A, 330, 79
\bibitem[\protect\citeauthoryear{Lyubarskii}{1999}]{l99}
Lyubarskii~Yu~E., 1999, MNRAS, 308, 1006
\bibitem[\protect\citeauthoryear{Lyutikov}{2003}]{Lyu03}
Lyutikov~M., 2003, MNRAS, 339, 623
\bibitem[\protect\citeauthoryear{McKinney, Tchekhovskoy \& Blanford}{2012}]{num3}
McKinney~J.C., Tchekhovskoy~A., Blanford~R.D., 2012, MNRAS, 423, 2083
\bibitem[\protect\citeauthoryear{McKinney \& Uzdensky}{2012}]{rec3}
McKinney~J.C.,  Uzdensky~D.A., 2012, MNRAS, 419, 573
\bibitem[\protect\citeauthoryear{Michel}{1969}]{Mich69}
Michel~F.C., 1969, ApJ, 158. 727
\bibitem[\protect\citeauthoryear{Nalewajko \& Begelman}{2012}]{NalBeg12}
Nalewajko~K., Begelman~M.C., 2012, MNRAS, 427, 2480
\bibitem[\protect\citeauthoryear{Nokhrina et al.}{2015}]{NBKZ}
Nokhrina~E.E., Beskin~V.S., Kovalev~Y.Y. Zheltoukhov~A.B., 2015, MNRAS, 447, 2726
\bibitem[\protect\citeauthoryear{Porth et al.}{2011}]{num2}
Porth~O., Fendt~Ch., Meliani~Z., Vaidya~B., 2011, ApJ, 737, 42
\bibitem[\protect\citeauthoryear{Reynolds et al}{1996}]{Rey96}
Reynolds~C.S., DiMatteo~T., Fabian~A.C., Hwang~U., Canizares~C.
1996 ,MNRAS, 283, L111
\bibitem[\protect\citeauthoryear{Romanova \& Lovelace}{1992}]{rec1}
Romanova~ M.M.,  Lovelace~ R.V.E., 1992, A\&A, 262, 26
\bibitem[\protect\citeauthoryear{Russo \& Thompson}{2013a}]{Russo1}
Russo~M., Thompson~Ch., 2013a, ApJ, 767, 142
\bibitem[\protect\citeauthoryear{Russo \& Thompson}{2013b}]{Russo2}
Russo~M., Thompson~Ch., 2013b, ApJ, 773, 99
\bibitem[\protect\citeauthoryear{Rybicki \& Lightman}{1981}]{Rybicki&Lightman}
Rybicki~G.B., Lightman~A.P., 1981, Radiative Processes in Astrophysics.
John Wiley \& Sons, New York
\bibitem[\protect\citeauthoryear{Sikora et al}{1996}]{Sikora1}
Sikora~M., Sol~H., Begelman~M.C, Madejski~G.M.,  1996, MNRAS, 280, 781
\bibitem[\protect\citeauthoryear{Sironi \& Spitkovsky}{2009}]{Sironi1}
Sironi~L., Spitkovsky~A., 2009, ApJ, 698, 1523
\bibitem[\protect\citeauthoryear{Stern \& Poutanen}{2006}]{SP06}
Stern~B.E., Poutanen~J., 2006, MNRAS, 372, 1217
\bibitem[\protect\citeauthoryear{Svensson}{1984}]{Sw}
Svensson~R., 1984, MNRAS, 209, 175
\bibitem[\protect\citeauthoryear{Takamoto \& Makoto}{2013}]{TaMa}
Takamoto, Makoto, 2013, ApJ, 775, 50T
\bibitem[\protect\citeauthoryear{Tchekhovskoy \& Bromberg}{2016}]{TchBro16}
Tchekhovskoy~A., Bromber~O, 2016, MNRAS, 461, L46
\bibitem[\protect\citeauthoryear{Tchekhovskoy et al.}{2008}]{Tch08}
Tchekhovskoy~A., McKinney~J., Narayan~R., 2008, MNRAS, 388, 551
\bibitem[\protect\citeauthoryear{Tchekhovskoy et al.}{2009}]{Tch09}
Tchekhovskoy~A., McKinney~J., Narayan~R., 2009, ApJ, 699, 1789
\bibitem[\protect\citeauthoryear{Thorne, Price \& Macdonald}{1986}]{TPM86}
Thorne~K.S., Price~R.H., Macdonald~D. 1986, Black Holes: The Membrane Paradigm.
Yale University Press, New Haven and London
\bibitem[\protect\citeauthoryear{Ustyugova et al}{1995}]{Ustyu95}
Ustyugova~G.V., Koldoba~A.V., Romanova~M.M., Che\-chet\-kin~V.M.,
Lovelace~R.V.E., 1995, ApJ, 439, L39
\bibitem[\protect\citeauthoryear{Del Zanna et~al.}{2016}]{DelZanna}
Del Zanna~L., Papini~E., Landi~S., Bugli~M., Bucciantini~N, 2016, arXiv:1605.06331
\end{thebibliography}
\end{document}